**Suppressed excitonic effects enable high-mobility, high-yield photoconductivity in a two-dimensional polymer crystal with axial pyridine coordination**

Shuai Fu[1,2,8]*, Ye Yang[1,7,8], Guoquan Gao[3,8], Shuangjie Zhao[1,8], Miroslav Položij[1,4,5], Tong Zhu[3], Lei Gao[2], Thomas Heine[1,4,5,6], Zhiyong Wang[1,7]*, Mischa Bonn[2]*, Xinliang Feng[1,7]*

[1]Center for Advancing Electronics Dresden and Faculty of Chemistry and Food Chemistry, Technische Universität Dresden, Dresden, Germany

[2]Max Planck Institute for Polymer Research, Mainz, Germany

[3]Laser Micro/Nano Fabrication Laboratory, School of Mechanical Engineering, Beijing Institute of Technology, Beijing, China

[4]Helmholtz-Zentrum Dresden-Rossendorf, HZDR, Dresden, Germany

[5]Center for Advanced Systems Understanding, CASUS, Görlitz, Germany

[6]Department of Chemistry and IBS Center for Nanomedicine, Yonsei University, Seoul, Republic of Korea

[7]Max Planck Institute of Microstructure Physics, Halle (Saale), Germany

[8]These authors contributed equally: Shuai Fu, Ye Yang, Guoquan Gao, Shuangjie Zhao

*E-Mail: shuai.fu@tu-dresden.de; zhiyong.wang@mpi-halle.mpg.de; bonn@mpip-mainz.mpg.de; xinliang.feng@mpi-halle.mpg.de

**Abstract**

Two-dimensional polymers (2DPs) and their layer-stacked covalent organic frameworks (2D COFs) offer modular, atomically precise platforms for organic optoelectronics, yet their photoconductive responses remain fundamentally constrained by strong excitonic effects and localized charge transport. Here, we demonstrate that a diyne-linked 2DP crystal with axial pyridine coordination overcomes this limitation, enabling simultaneous efficient free-carrier generation and band-like transport. Introducing pyridine ligands that axially coordinate to Cu–porphyrin nodes transforms weak van der Waals stacking into a pyridine-bridged architecture with pronounced interlayer band dispersion and substantially reduced carrier effective masses. The resulting strong interlayer electronic coupling suppresses the exciton binding energy to well below thermal energy, such that optical excitation directly populates delocalized electronic states. Time-resolved terahertz spectroscopy reveals Drude-type photoconductivity with room-temperature mobilities approaching 500 $cm^2 \cdot V^{-1} \cdot s^{-1}$ and a photon-to-free-carrier conversion ratio of ~0.4, yielding a photoconductive response that exceeds that of state-of-the-art organic and many inorganic photoactive materials. These results establish interlayer coordination as a powerful strategy for mitigating excitonic effects and accessing inorganic-like charge transport in organic 2D crystals, opening a pathway toward highly efficient photo-to-electricity conversion in organic-based systems.

**Main text**

Organic semiconductors combine strong light–matter interaction, synthetic versatility, and mechanical flexibility, and have therefore become central to emerging optoelectronic technologies[1-3]. Their modular molecular architectures enable the energy landscape and charge transport pathways to be engineered with atomic precision[4-6]. The advent of organic two-dimensional crystals (O2DCs), including crystalline 2D polymers (2DPs) and their layer-stacked covalent organic frameworks (2D COFs), as well as conjugated metal-organic frameworks (2D *c*-MOFs), has further transformed the field[7-13]. Long-range order and periodic framework connectivity in these materials mitigate energetic and structural disorder, giving rise to dispersive electronic bands and room-temperature mobilities that can approach or exceed those of benchmark organic semiconductors and, in some cases, rival those of inorganic systems over short length scales[14]. In terms of charge transport alone, O2DCs have thus largely overcome the limitations that have long constrained conventional conjugated polymers.

For light–to–electricity conversion, however, high mobility is necessary but not sufficient. Optical excitation in organic semiconductors predominantly generates tightly bound Frenkel excitons, reflecting localized electronic wavefunctions and weak dielectric screening; consequently, the exciton binding energy ($E_b$) is typically orders of magnitude larger than in inorganic semiconductors[15,16]. Under practical excitation conditions, optical absorption therefore produces bound electron–hole pairs rather than free carriers, and efficient operation of organic photovoltaics and photodetectors relies on donor–acceptor heterojunctions or strong internal electric fields to drive exciton dissociation[17,18]. Even in high-mobility O2DCs, reported photoconductivity yields remain substantially below those of state-of-the-art inorganic photoactive materials[14,19,20], indicating that the photon-to-free-carrier conversion ratio ($\phi$) remains the primary bottleneck (**Fig. 1a**). O2DCs can thus transport charges efficiently when present, yet still fail to generate them with the efficiency required for truly competitive photovoltaics and photodetection.

The relevant figure of merit is the photoconductivity $\Delta\sigma = n \cdot e \cdot \mu$, where $n$ is the density of photogenerated free carriers, $e$ is the elementary charge, and $\mu$ is the charge mobility. Considerable effort has been devoted to maximizing $\mu$ in O2DCs through backbone design[21,22], substitution patterning[23], and suppression of grain-boundary scattering[19], yielding impressive band-like transport characteristics[14,24]. By contrast, strategies to increase $n$ by suppressing excitonic effects have largely focused on intralayer modifications, including donor–acceptor motifs[25,26], frontier-orbital and bandgap engineering[27], control of backbone planarity[28,29], and the introduction of π-bridges[30]. While these approaches can reduce $E_b$ into the few-tens-of-meV regime[31], they generally operate within individual layers and often introduce additional structural complexity or compromise charge transport. Prior seminal work[32] has shown that extending π-conjugation across 2D sheets can substantially increase $\phi$; however, the potential of interlayer electronic coupling as an independent design parameter for controlling exciton binding and free-carrier generation in O2DCs has remained comparatively unexplored.

Here, we introduce a complementary design concept that targets $E_b$ through engineered interlayer electronic coupling via interlayer coordination, rather than relying solely on intralayer structural modulation. We find that converting weak van der Waals stacking into a coordination-bridged, electronically coherent architecture can delocalize the exciton wavefunction in three dimensions, reduce effective masses, and enhance dielectric screening, thereby pushing an intrinsically organic crystal toward Wannier-like behavior in which $E_b \ll k_B T$ and optical excitation directly populates delocalized electronic states. We employ a diyne-linked, AB-stacked Cu–porphyrin 2DP with intrinsic interlayer voids and accessible axial coordination sites, which the theoretical calculations predicted to host pyridine ligands as periodic coordination bridges between adjacent layers. As we show below, this interlayer coordination motif simultaneously suppresses excitonic effects and preserves band-like transport, yielding a single-component 2DP crystal in which light absorption produces Drude-type free carriers with high mobility (~500 cm$^2$·V$^{-1}$·s$^{-1}$) and large photoconductivity yield (~0.4), thereby directly addressing the long-standing exciton bottleneck that has limited the optoelectronic performance of organic semiconductors.

**On-liquid surface synthesis of PI-DY2DP with axial pyridine coordination**

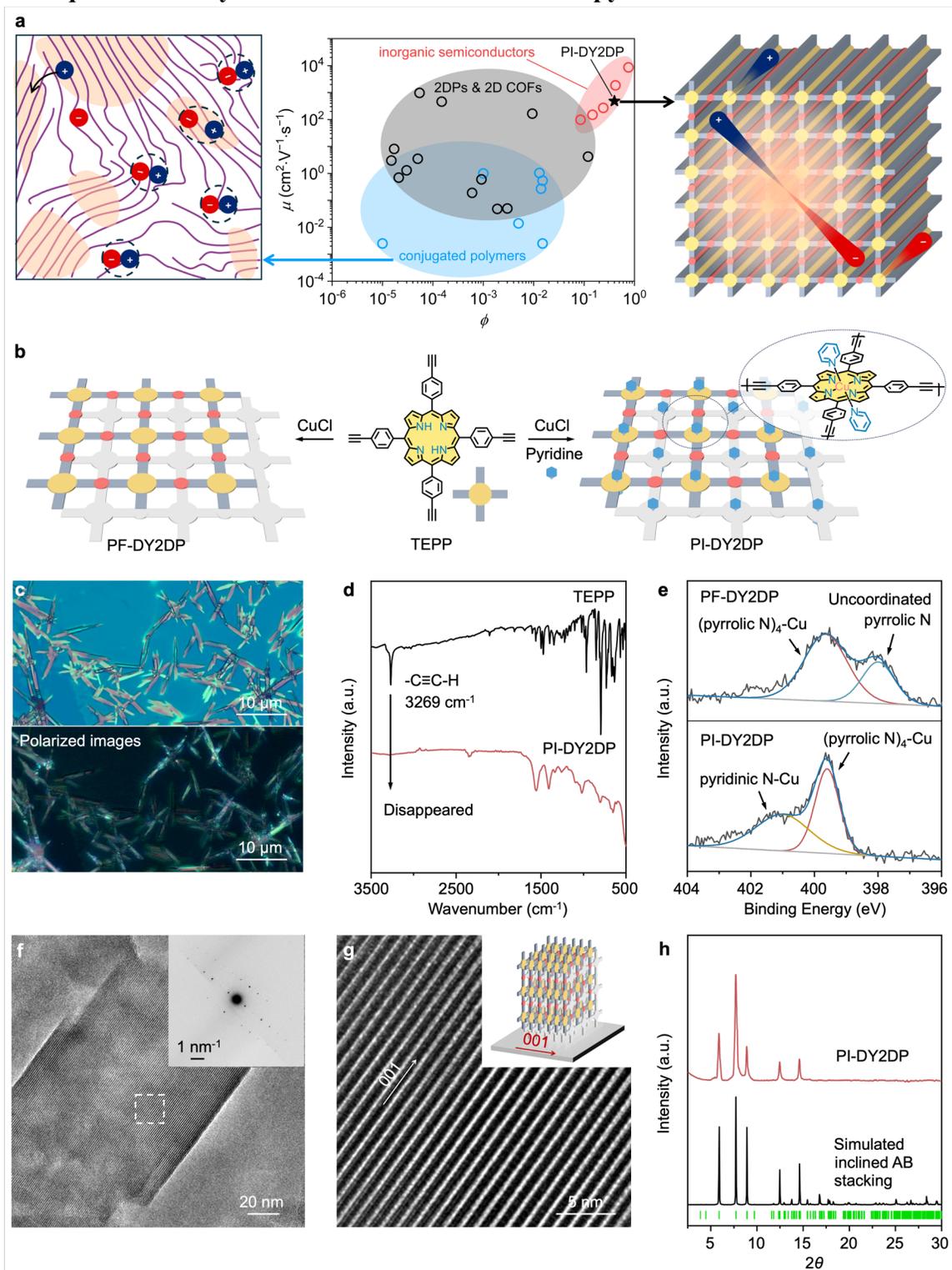

**Fig. 1 | Synthesis and structural characterization of DY2DPs. a,** Comparison of $\mu$ and $\phi$ across representative material classes: conjugated polymers, 2DPs and 2D COFs, and inorganic semiconductors. The left panel schematically illustrates photoexcitation in

conjugated polymers with relatively high $E_b$, where optical excitation predominantly produces localized and/or weakly delocalized excitons, with only a small fraction of dissociated carriers undergoing hopping transport across semicrystalline domains. The right panel depicts photoexcitation in certain 2DPs, 2D COFs, and inorganic semiconductors with relatively low $E_b$, where excitation directly generates free carriers that undergo band-like transport within well-ordered lattices. **b,** Synthetic schemes for PF-DY2DP and PI-DY2DP. **c,** OM (top) and polarized OM (bottom) images of PI-DY2DP. **d,** ATR-FTIR spectra of the TEPP monomer and PI-DY2DP. **e,** High-resolution XPS N 1s spectra of PF-DY2DP (top) and PI-DY2DP (bottom). **f,** HRTEM image of PI-DY2DP; insert shows the corresponding SAED pattern. **g,** Magnified HRTEM image of PI-DY2DP; insert schematically illustrates the inclined AB stacking and the preferential edge-on orientation of PI-DY2DP on the substrate. **h,** Experimental XRD pattern of PI-DY2DP and simulated pattern based on the inclined AB-stacked structural model.

Theoretical simulations showed that the intrinsic voids in the AB-stacked Cu–porphyrin 2DP structure can accommodate small molecules, which can act as axial ligand bridges between the Cu centers. Pyridine was selected as the axial ligand owing to its unique combination of moderate coordination strength (interaction energy, –67 kJ·mol$^{-1}$), structural simplicity, and electronic compatibility with π-conjugated frameworks[33,34]. As a neutral σ-donor and weak π-acceptor[35-37], pyridine forms stable yet reversible axial bonds with Cu–porphyrins[38], enabling well-defined coordination without disrupting the in-plane conjugation of the 2DP lattice (Supplementary Fig. 1). Importantly, its small molecular size and planar aromatic character permit dense, periodic interlayer coordination, thereby maximizing orbital overlap between adjacent layers while minimizing steric distortion. This balance is expected to strengthen interlayer electronic coupling and facilitate interlayer charge transport. To establish a comparative platform for probing axial pyridine coordination effects, we synthesized both pyridine-free DY2DPs (PF-DY2DPs) and PI-DY2DPs using a three-step on-organic-liquid surfactant-monolayer-assisted interfacial synthesis (O-SMAIS) strategy (**Fig. 1b** and Supplementary Fig. 2)[39,40]. In Step I, a perfluoro-surfactant (PFS) monolayer was prepared on the N,N-dimethylacetamide/water (DMAc)-H$_2$O surface, followed by addition of an aqueous CuCl solution (3 × 10$^{-5}$ mol), with or without pyridine (6 × 10$^{-5}$ mol), to generate a catalyst-enriched interfacial region via electrostatic accumulation of Cu$^+$ ions beneath the anionic surfactant (Step II). Subsequently, 5,10,15,20-tetrakis(4-ethynylphenyl)porphyrin (TEPP) monomers (4.2 × 10$^{-6}$ mol) in DMAc solution were injected into the subphase, where coordination between terminal alkynes and Cu$^+$ induced vertical assembly and initiated 2D polymerization. After

24 h, the resulting DY2DPs were transferred onto solid substrates (e.g., $SiO_2$/Si wafers, transmission electron microscopy (TEM) grids, or fused silica) by horizontal lifting for morphological, structural, or property characterization.

Polarized optical microscopy (OM) reveals that PI-DY2DP forms rod-shaped crystals with typical lengths of ~10 μm and widths of ~0.5 μm (**Fig. 1c**), whereas PF-DY2DP forms nanocrystals with typical lateral dimensions of ~0.3 × 0.3 μm² (Supplementary Fig. 3), highlighting a pronounced pyridine-induced modulation of crystal growth. Attenuated total reflection Fourier-transform infrared (ATR-FTIR) spectra show complete disappearance of the C≡C-H stretching vibration of TEPP at 3269 $cm^{-1}$ after polymerization, confirming full conversion of terminal alkynes into diyne linkages (**Fig. 1d**). The formation of diyne linkages, rather than possible Cu-C bonds, is attributed to enhanced alkyne reactivity at the $Cu^+$-rich interface relative to the bulk solution, and is further supported by a model reaction (Supplementary Fig. 4). X-ray photoelectron spectroscopy (XPS) provides direct evidence of axial pyridine coordination. Deconvolution of the N 1s spectrum of PF-DY2DP reveals two components assigned to (pyrrolic N)$_4$-Cu (399.6 eV) and uncoordinated pyrrolic N (398.0 eV) (**Fig. 1e**, top). By contrast, PI-DY2DP exhibits an additional peak at 401.1 eV, attributable to pyridinic N-Cu coordination (**Fig. 1e**, bottom), unambiguously confirming incorporation of axially coordinated pyridine. Furthermore, the ratio of pyrrolic N-Cu to pyridinic N-Cu species (2:1.3) closely approaches the theoretical 2:1 value expected for Cu centers coordinated by two axial pyridine ligands. High-resolution transmission electron microscopy (HRTEM) of PI-DY2DP resolves periodic lattice fringes with $d$-spacings of 11.5 and 7.2 Å (**Fig. 1f, g**). Consistently, selected-area electron diffraction (SAED) patterns display multiple sets of reflections, with the innermost spots at 0.88 and 0.14 $nm^{-1}$ corresponding to the (200) and (001) crystallographic planes, respectively (**Fig. 1f**, insert). X-ray diffraction (XRD) analysis reveals sharp Bragg peaks at $2\theta$ = 5.88°, 7.71°, 8.89°, 12.47°, and 14.60°, indexed to the (110), (200), (020), (021), and (212) planes, respectively (**Fig. 1h**). These data define a square unit cell with lattice parameters $a = b = 22.9$ Å and $c = 7.3$ Å ($\alpha = \gamma = 90°$, $\beta = 60°$), corresponding well with the inclined AB-stacking model. HRTEM of PF-DY2DP displays lattice fringes with a $d$-spacing of ~1.1 nm (Supplementary Fig. 5), comparable to those observed for PI-DY2DP, while XRD patterns exhibit nearly identical diffraction features (Supplementary Fig. 6; $2\theta$ = 5.89°, 7.69°, 8.89°,

12.51°, and 14.62°), indicating that both materials share the same lattice framework, and that the principal distinction resides in their interlayer chemical motifs. In PI-DY2DP, axial coordination of pyridine to the Cu centers within the porphyrin units reinforces interlayer registry and establishes periodic coordination bridges between adjacent layers (Supplementary Fig. 1).

**Giant photoconductive response enabled by coordination-induced interlayer coupling**

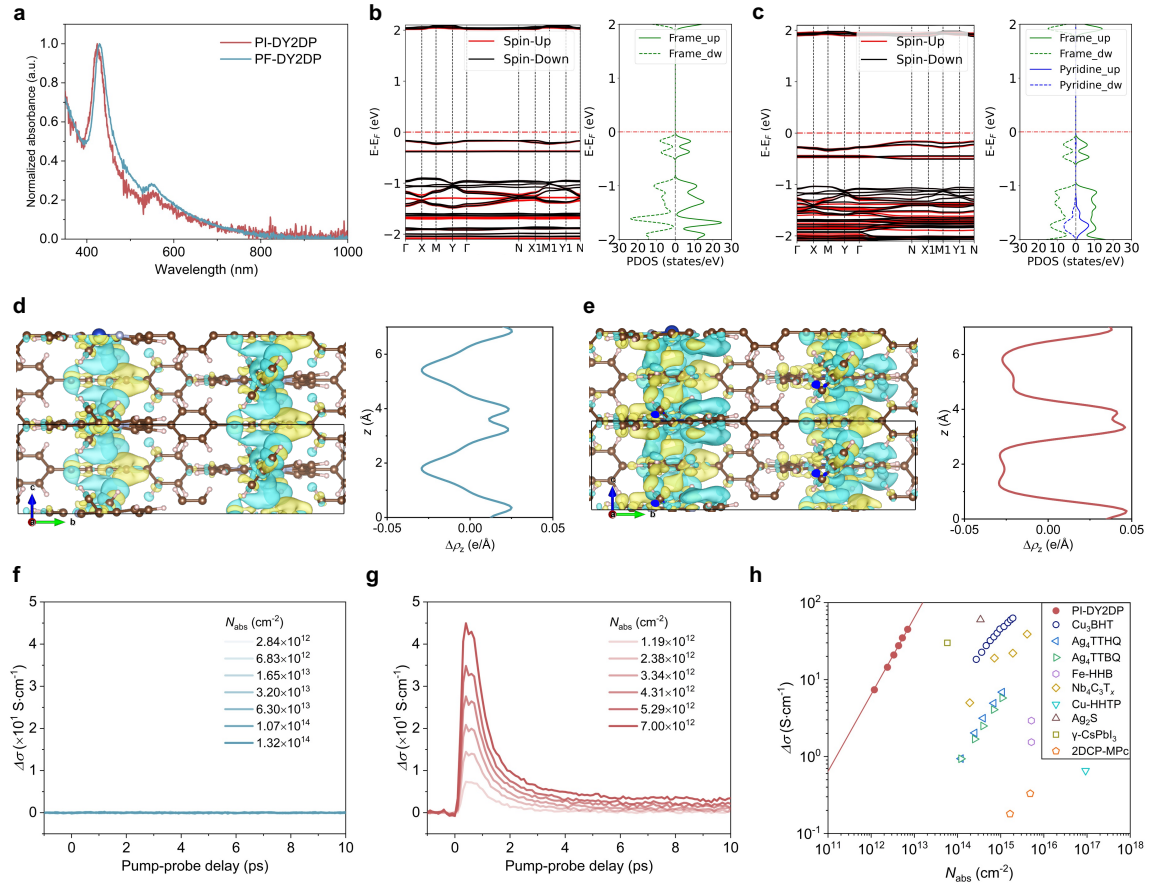

**Fig. 2 | Axial pyridine coordination reshapes the electronic landscape and enables giant photoconductive response. a,** UV–vis absorption spectra of PF-DY2DP and PI-DY2DP. **b,c,** Calculated electronic band structures and PDOS of PF-DY2DP (**b**) and PI-DY2DP (**c**). The Brillouin-zone high-symmetry $k$-path is shown in Supplementary Fig. 7. **d,e,** Side view of CDD maps and plane-averaged differential charge density profiles of PF-DY2DP (**d**) and PI-DY2DP (**e**). Yellow and cyan denote charge accumulation and depletion, respectively; black lines mark the unit-cell boundary. **f,** Time-resolved $\Delta\sigma$ of PF-DY2DP, showing no detectable signal within the noise level over an $N_{abs}$ range of 2.84 × $10^{12}$ to 1.32 × $10^{14}$ cm$^{-2}$. **g,** Time-resolved $\Delta\sigma$ of PI-DY2DP measured over $N_{abs}$ = 1.19 × $10^{12}$ to 7.00 × $10^{12}$ cm$^{-2}$. **h,** $\Delta\sigma_{peak}$ as a function of $N_{abs}$ for PI-DY2DP compared with representative organic and inorganic photoactive materials characterized by TRTS.

To examine how axial pyridine coordination reshapes the electronic landscape of DY2DPs, we first compared the optical absorption spectra of PI-DY2DP and PF-DY2DP. As shown in **Fig. 2a**, both materials exhibit nearly identical UV–vis absorption profiles, with two dominant peaks at 2.24 and 2.91 eV and comparable optical bandgaps ($E_{opt}$) of 2.04 eV extracted from Tauc plots (Supplementary Fig. 8). This corresponds well with the electronic band structure and projected density of states (PDOS) calculations obtained with the HSE06 hybrid functional (**Fig. 2b, c**; Supplementary Fig. 1). Both PF-DY2DP and PI-DY2DP are semiconductors with similar intralayer (i.e., X–M–Y) band dispersions. In striking contrast, their interlayer (i.e., Γ–N) band dispersions differ profoundly: PF-DY2DP exhibits nearly flat conduction and valence bands near the Fermi level, indicative of weak interlayer orbital overlap and correspondingly large interlayer electron and hole masses exceeding 10 $m_0$ (the precise values cannot be reliably determined because of the vanishing band curvature). By comparison, PI-DY2DP shows pronounced Γ–N dispersion, yielding markedly reduced interlayer electron and hole masses of 0.74 $m_0$ and 0.24 $m_0$, respectively. These results demonstrate that axial pyridine coordination activates strong interlayer electronic coupling, thereby facilitating efficient charge transport along the stacking direction. This conclusion is further supported by charge-density-difference (CDD) analysis, which reveals enhanced electronic delocalization across adjacent layers in PI-DY2DP relative to PF-DY2DP. In PF-DY2DP, charge redistribution is largely confined to the organic linker regions, arising from weak interlayer π-π interactions, with minimal charge rearrangement at the Cu-porphyrin centers (**Fig. 2d** and Supplementary Fig. 9a). By contrast, PI-DY2DP exhibits pronounced charge delocalization along the Cu–N coordination axis (**Fig. 2e** and Supplementary Fig. 9b), establishing a continuous electronic pathway bridging adjacent layers, consistent with its enhanced interlayer band dispersion and markedly reduced carrier effective masses. Projected density of states (PDOS) analysis further reveals that the band-edge states are dominated by the DY2DP backbone, whereas pyridine contributes primarily to deeper valence-band states, indicating that pyridine acts as a coordination-mediated electronic bridge rather than as a band-edge contributor. The calculated electrical bandgaps ($E_{elec}$) of PI-DY2DP and PF-DY2DP are 2.04 eV and 2.17 eV, respectively. Using $E_b=E_{elec}-E_{opt}$, we estimate an $E_b$ value for PI-DY2DP that is negligibly small relative to thermal energy at room temperature (~25.7 meV), whereas PF-

DY2DP exhibits a substantially larger $E_b$ of ~130 meV. The ultralow $E_b$ of PI-DY2DP indicates that axial pyridine coordination facilitates efficient exciton dissociation, thereby providing an energetically favorable landscape for free-carrier generation and photo-to-electricity conversion.

Next, we employed time-resolved terahertz spectroscopy (TRTS) to directly probe the dynamics of photogenerated charge species in DY2DPs. In a typical TRTS experiment, an ultrashort pump pulse (~100 fs, 3.1 eV) excites interband transitions, generating excitonic and/or free carrier populations. A time-delayed, single-cycle terahertz (THz) probe pulse (~1 ps duration) subsequently traverses the sample, resolving the transient $\Delta\sigma$ in both the time and frequency domains. This provides direct access to the nature and intrinsic transport properties of photogenerated charge species[41,42]. Upon photoexcitation, PF-DY2DP exhibits no detectable $\Delta\sigma$ signal (< 0.1 S·cm$^{-1}$) over absorbed photon densities ($N_{abs}$) up to $1.32 \times 10^{14}$ cm$^{-2}$ (**Fig. 2f**). The absence of measurable photoconductivity in PF-DY2DP suggests that the photoinduced species predominantly remain as tightly bound Frenkel excitons that do not contribute to the real part of $\Delta\sigma$, or that any transiently dissociated carriers rapidly recombine or experience spatial confinement within small crystalline domains. This behavior is consistent with its relatively large $E_b$, nearly flat interlayer electronic bands, and lower crystallinity, all of which favor charge localization. In striking contrast, PI-DY2DP displays a pronounced rise in $\Delta\sigma$ immediately after photoexcitation, even at $N_{abs}$ values two orders of magnitude lower (down to $1.19 \times 10^{12}$ cm$^{-2}$, **Fig. 2g**). This photoconductivity response is highly reproducible across multiple spatial positions and independent synthesis batches (variations < 10%, Supplementary Fig. 10). This dramatic contrast in photoconductivity yield—approaching ~10$^4$-fold—underscores the critical role of axial pyridine coordination in activating high photoconductive response in DY2DP; while contributions from crystal dimensions and morphology cannot be fully excluded due to the intrinsic difficulty in obtaining PF-DY2DP samples of comparable quality. Using $\Delta\sigma = N_{abs} \cdot \phi \cdot e \cdot \mu$, we extract an effective mobility ($\phi \cdot \mu$) of 200 cm$^2$·V$^{-1}$·s$^{-1}$ for PI-DY2DP, which represents a lower bound for $\mu$ (since $0 < \phi < 100\%$). This large effective mobility directly evidences the concurrent realization of efficient free-carrier generation and rapid charge migration in PI-DY2DP, consistent with the coordination-induced reduction of both $E_b$ and $m^*$. Further insight is obtained from

$N_{abs}$-dependent measurements. The peak photoconductivity ($\Delta\sigma_{peak}$) increases linearly from 7.4 to 44.9 S·cm$^{-1}$ as $N_{abs}$ rises from $1.19 \times 10^{12}$ to $7.00 \times 10^{12}$ cm$^{-2}$ (Supplementary Fig. 11a), while the decay dynamics remain essentially unchanged (Supplementary Fig. 11b). This linear scaling and $N_{abs}$-independent decay rate indicate that free carriers originate from direct population of delocalized electronic states in this excitation regime, rather than from nonlinear processes such as singlet–singlet annihilation, which would produce sublinear $\Delta\sigma_{peak}$-$N_{abs}$ dependence accompanied by accelerated decay at higher excitation densities[43]. Notably, the large $\Delta\sigma$ achieved even under low excitation densities corresponds to an exceptional photoconductivity yield of $6.4 \times 10^{-12}$ S·cm, surpassing that of state-of-the-art organic and many inorganic photoactive materials measured using the same methodology (**Fig. 2h**), including conjugated coordination polymers (Cu$_3$BHT, Ag$_4$TTHQ, Ag$_4$TTBQ, Fe-HHB, Cu-HHTP)[44-47], 2D COFs (2DCP-MPc)[14], and inorganic semiconductors (Ag$_2$S, and CsPbI$_3$)[48,49].

**Temporal evolution of Drude-type, free-carrier dominated photoconductivity**

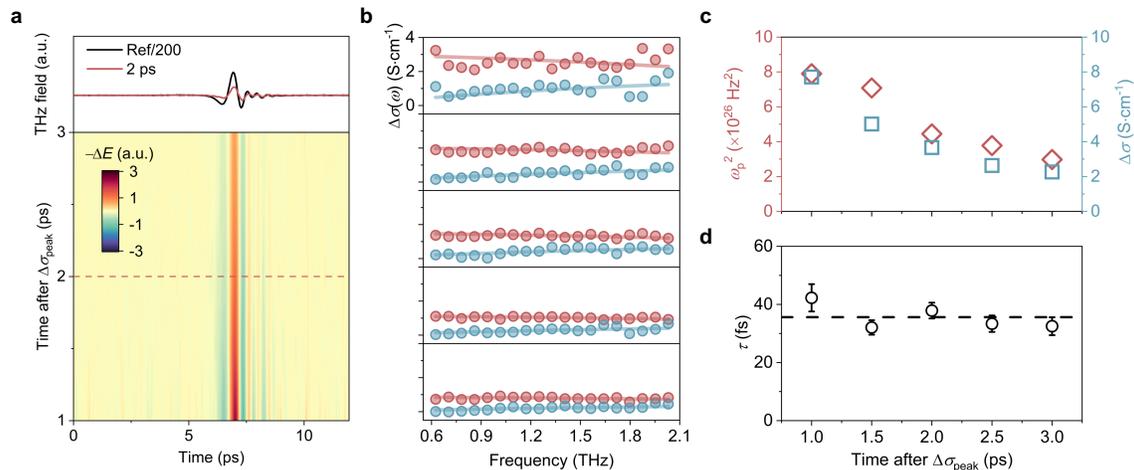

**Fig. 3 | Temporal evaluation of THz photoconductivity governed by Drude–type free carriers in PI-DY2DP. a,** Top: Time–resolved THz electric field transmitted through unexcited PI-DY2DP (scaled by a factor of 200) and pump–induced changes at a representative delay (2 ps after $\Delta\sigma_{peak}$). Bottom: Pseudo–color map of pump–induced THz electric field changes as a function of pump–probe delay relative to $\Delta\sigma_{peak}$. Measurements were performed under 3.10 eV photoexcitation at $N_{abs} = 3.34 \times 10^{12}$ cm$^{-2}$ at room temperature. **b,** Frequency–resolved complex THz photoconductivity, $\Delta\sigma(\omega)$, at selected delays after $\Delta\sigma_{peak}$. Red and blue symbols denote the real and imaginary components of $\Delta\sigma(\omega)$, respectively, and solid lines show the corresponding Drude fits. **c,** Temporal

evolution of $\omega_\text{p}^2$ (red diamonds, left axis) and $\Delta\sigma$ (blue squares, right axis). **d,** Temporal evolution of $\tau$.

To uncover the microscopic origin of the exceptional photoconductive response observed in PI-DY2DP, we analyzed time-resolved THz waveforms recorded at various pump–probe delays following $\Delta\sigma_\text{peak}$ (**Fig. 3a**), from which the frequency-resolved complex photoconductivity $\Delta\sigma(\omega)$ was extracted (Methods). The $\Delta\sigma(\omega)$ spectra measured at different time delays exhibit nearly identical frequency dispersions (**Fig. 3b**), characterized by a finite positive real component and an imaginary component close to zero that gradually approaches the real part at higher frequencies. This spectral signature is characteristic of delocalized free-carrier transport with a momentum scattering rate exceeding the spectral bandwidth of the THz probe, and can be well described by the Drude model:

$$\Delta\sigma(\omega) = \frac{ne^2\tau}{m^*(1-i\omega\tau)}, n = \frac{\omega_\text{p}^2 m^* \varepsilon_0}{e^2}$$

where $\tau$ is the momentum-averaged carrier scattering time, $\omega_\text{p}$ is the plasma frequency, and $\varepsilon_0$ is the vacuum permittivity. Fitting $\Delta\sigma(\omega)$ with the Drude model enables direct tracking of the temporal evolution of the microscopic transport parameters and disentangles the respective contributions of $\tau$ and $\omega_\text{p}^2$ to $\Delta\sigma$. As shown in **Fig. 3c**, $\omega_\text{p}^2$—proportional to the free-carrier density ($n$)—decays in concert with $\Delta\sigma$, whereas $\tau$ remains essentially constant at 36 ± 4 fs throughout the measured time window. This time-invariant $\tau$ indicates that attenuation of $\Delta\sigma$ is governed primarily by decay of the free-carrier population rather than by evolving carrier localization or carrier-carrier interactions, consistent with transport in a dilute free-carrier regime. Using the momentum-averaged reduced electron-hole effective mass $m^* = 0.13$ m$_0$ obtained from DFT calculations (Supplementary Table 1), the room-temperature d.c. mobility $\mu$, estimated via $\mu = e\tau/m^*$, reaches 480 ± 50 cm$^2\cdot$V$^{-1}\cdot$s$^{-1}$ for PI-DY2DP, placing it among the highest mobilities reported for 2DPs and 2D COFs (Supplementary Table 2). Combining $\mu$ with the experimentally extracted effective mobility yields a $\phi$ vale of 42%, substantially exceeding that of organic materials measured by TRTS or flash-photolysis time-resolved microwave conductivity combined with the current integration method (Supplementary Table 3). These results demonstrate that axial pyridine coordination substantially suppresses excitonic effects, enabling the rare convergence of efficient free-carrier generation and high-mobility transport in organic materials.

**Charge transport and recombination mechanisms in PI-DY2DP**

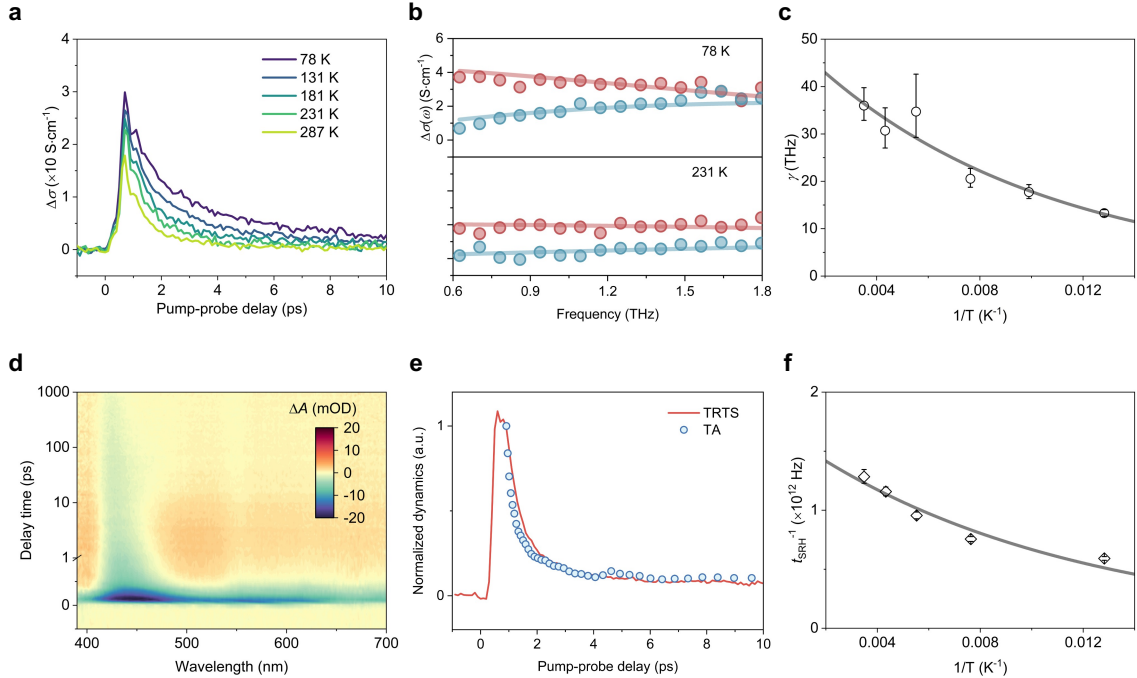

**Fig. 4 | Band–like transport and SRH recombination in PI-DY2DP. a,** Temperature–dependent time-resolved $\Delta\sigma$ under 3.10 eV photoexcitation at $N_{abs} = 2.75 \times 10^{12}$ cm$^{-2}$. **b,** Frequency-resolved complex THz photoconductivity, $\Delta\sigma(\omega)$, measured at 2 ps after $\sigma_{peak}$ at 78 K (top) and 231 K (bottom). Red and blue symbols denote the real and imaginary components of $\Delta\sigma(\omega)$, respectively, and solid lines represent the corresponding Drude fits. **c,** Temperature dependence of the charge scattering rate. Circles and error bars indicate mean values and standard errors extracted from Drude fits to the temperature–dependent $\Delta\sigma(\omega)$ spectra, respectively (Supplementary Fig. 12). The solid line shows an Arrhenius fit. **d,** Pseudo–color 2D TA map as a function of probe wavelength and pump–probe delay under 3.10 eV photoexcitation at $N_{abs} = 2.9 \times 10^{12}$ cm$^{-2}$. **e,** Comparison of normalized carrier dynamics extracted from TRTS (red line) and TA (blue dots). **f,** Temperature dependence of the characteristic time constants associated with SRH recombination. Diamonds and error bars denote mean values and standard errors obtained from bi–exponential fits to the dynamics in panel **a**. The solid line represents a fit based on the SRH recombination model.

To elucidate the charge transport mechanism in PI-DY2DP, we conducted temperature-dependent TRTS experiments. **Fig. 4a** shows the time-resolved $\Delta\sigma$ at different temperatures, revealing systematic changes in both amplitude and decay dynamics. To understand the temperature evolution of $\Delta\sigma$, we extracted $\Delta\sigma(\omega)$ at each temperature (**Fig.**

**4b** and Supplementary Fig. 12). All $\Delta\sigma(\omega)$ spectra remain well described by the Drude model across the entire temperature range used, indicating that the free-carrier character persists upon cooling. Notably, the crossover frequency between the real and imaginary components of $\Delta\sigma(\omega)$ shifts to lower frequencies as temperature decreases, corresponding to an increase in $\tau$ from 28 ± 3 fs at 287 K to 75 ± 4 fs at 78 K. The resulting negative temperature coefficient of $\tau$ (i.e., $d\tau/dT < 0$) is characteristic of phonon-limited transport and represents a hallmark of band-like charge transport. Arrhenius analysis of the scattering rate $\gamma$ ($\gamma = 1/\tau$) yields an activation energy of approximately 10 meV, suggesting that carrier scattering is governed by a low-energy phonon mode of comparable energy (**Fig. 4c**).

To clarify the origin of the picosecond decay of $\Delta\sigma$ and its temperature dependence, we performed complementary transient absorption (TA) measurements. While TRTS selectively tracks mobile free carriers, TA is sensitive to both free carriers and bound excitons, enabling discrimination between free-carrier recombination and exciton formation. **Fig. 4d** presents a pseudo-color TA map under 3.10 eV photoexcitation, revealing two prominent photoinduced bleach (PIB) features at 2.25 eV (~550 nm) and 2.95 eV (~420 nm) that coincide with steady-state absorption peaks. Comparison of normalized TRTS and TA dynamics reveals closely matched decay profiles on the picosecond timescale (**Fig. 4e**). This correspondence indicates that the initial drop in $\Delta\sigma$ originates predominantly from ultrafast non-radiative free-carrier recombination, rather than conversion into long-lived excitons, which would be expected to produce divergent TRTS and TA dynamics[50]. Analysis of the temperature dependence of the fast decay component further supports this interpretation. The associated time constant increases upon cooling (**Fig. 4f**), consistent with Shockley–Read–Hall (SRH) recombination, in which free carriers are captured by trap states prior to non-radiative recombination[51]. Fitting the temperature-dependent decay constants with an SRH model yields a trap depth of ~8 meV, indicating that shallow defect states dominate the recombination process. The identification of shallow trap states as the primary recombination centers highlights clear opportunities for defect engineering—through chemical passivation, improved structural order, or electrostatic trap filling—to further extend free-carrier lifetimes, diffusion lengths, and device-relevant performance.

**Outlook**

Single-component organic semiconductors typically exhibit low photoconductive responses because they lack both sufficient thermodynamic driving force to overcome strong excitonic binding and extended electronic states that support band-like transport. Consequently, photoexcitation predominantly generates tightly bound Frenkel excitons and, at best, partially dissociated free carriers that hop through an energetically disordered landscape. This limitation often necessitates donor–acceptor architectures, interfacial charge separation, excess kinetic energy, and/or strong internal electric fields to achieve efficient photon-to-electricity conversion in practical applications. The recent emergence of O2DCs has opened new opportunities by providing high-mobility charge transport pathways. However, the full potential of photoconductive response optimization can only be realized if the pronounced excitonic effects arising from weak van der Waals interactions between adjacent layers are effectively suppressed.

Here, we exploit the intrinsic interlayer voids of an AB-stacked DY2DP to introduce axial pyridine coordination, transforming weak van der Waals stacking into a pyridine-bridged architecture. This coordination-induced interlayer coupling substantially suppresses excitonic effects and enhances interlayer band dispersion, enabling the simultaneous realization of efficient free-carrier generation and band-like transport. These synergistic effects produce an exceptional photoconductive response comparable to that of benchmark inorganic semiconductors. More broadly, this work establishes a molecular-level strategy for reconciling exciton dissociation and charge transport in organic semiconductors. It highlights that interlayer interactions can be as decisive as intralayer structural modulation in shaping the electronic landscape of O2DCs, enabling photoconductive characteristics traditionally associated with only leading inorganic photoactive materials. Future advances may arise from diversifying interlayer interaction motifs, precisely tuning coupling strength, and mitigating shallow trap states through targeted defect engineering. Together, these directions point toward the rational design of O2DCs for high-efficiency photo-to-electricity conversion, optoelectronic devices, and photocatalytic systems.

**Acknowledgements**

This work was financially supported by the ERC Synergy Grant (2DPolyMembrane, grant no. 101167472), ERC Consolidator Grant (T2DCP), GRK2861 (grant no. 491865171), CRC 1415 (Chemistry of Synthetic Two-Dimensional Materials, no. 417590517), the DFG Priority Program SPP 2244 "2DMP" (project no. 443405902), as well as the German Science Council and Center of Advancing Electronics Dresden. The authors acknowledge Dr. Darius Pohl and Dr. Bernd Rellinghaus for providing access to the TEM facilities at the Dresden Center for Nanoanalysis.

**Author contributions**

S.F., Z.W., M.B. and X.F. conceived the project. Y.Y. and S.F. contributed to the synthesis, routine characterization, and data analysis under the supervision of Z.W. and X.F. S.F. and L.G. performed the TRTS measurements and analyzed the results under the supervision of M.B. G.G. and T.Z. conducted the TA measurements. S.Z., M.P. and T.H. performed the theoretical calculations and determined the interlayer coordination of pyridines. S.F., Y.Y. and Z.W. co-wrote the manuscript with input from all authors. All authors discussed the results and commented on the manuscript.

**Competing interests**

The authors declare no competing financial interest.

**Methods**

**On-liquid surface synthesis of PI-DY2DPs.** A 40 mL mixture of DMAc and Milli-Q water (v/v = 1:1) was transferred into a crystallization dish (6 cm in diameter). Subsequently, 28 μL of a PFS solution (1 mg mL$^{-1}$ in chloroform) was carefully spread onto the DMAc-$H_2O$ interface using a micropipette to form a surfactant monolayer. After allowing the monolayer to stabilize for 30 min, an aqueous solution containing CuCl (0.03 mmol) and pyridine (0.06 mmol; total volume 1 mL) was gently injected into the aqueous subphase with a syringe. Following an incubation period of 2 h, TEPP (4.2 μmol in DMAc) was introduced into the system. The reaction was then allowed to proceed under quiescent conditions at 1 °C for 24 h. Subsequently, rod-shaped crystals formed at the interface and were transferred onto substrates via a horizontal dipping method. The resulting products were thoroughly rinsed with Milli-Q water and ethanol, and finally dried under a nitrogen flow.

**On-liquid surface synthesis of PF-DY2DPs.** A 40 mL mixture of DMAc and Milli-Q water (v/v = 1:1) was transferred into a crystallization dish (6 cm in diameter). Subsequently, 28 μL of a PFS solution (1 mg mL$^{-1}$ in chloroform) was carefully spread onto the DMAc-$H_2O$ interface using a micropipette to form a surfactant monolayer. After allowing the monolayer to stabilize for 30 min, an aqueous solution containing CuCl (0.03 mmol; total volume 1 mL) was gently injected into the aqueous subphase with a syringe. Following an incubation period of 2 h, TEPP (4.2 μmol in DMAc) was introduced into the system. The reaction was then allowed to proceed under quiescent conditions at 1 °C for 14 days. Subsequently, nanocrystals formed at the interface and were transferred onto substrates via a horizontal dipping method. The resulting products were thoroughly rinsed with Milli-Q water and ethanol, and finally dried under a nitrogen flow.

**Time-resolved terahertz spectroscopy (TRTS).** Time-resolved terahertz spectroscopy was employed to probe the ultrafast photoconductive response of the samples in both the time and frequency domains. The measurements were carried out using a femtosecond laser system based on a mode-locked Ti:sapphire regenerative amplifier, delivering pulses centered at 1.55 eV with a pulse width of ~50 fs and a repetition rate of 1 kHz. Photoexcitation at 3.10 eV was obtained via second-harmonic generation of the fundamental output using a $\beta$-$BiB_3O_6$ nonlinear crystal. Single-cycle THz pulses (~1 ps duration) were generated by optical rectification of the 1.55 eV beam in a 1-mm-thick (110)-oriented ZnTe crystal and detected in a second ZnTe crystal through free-space electro-optic sampling. The emitted THz radiation was guided to the sample in a transmission geometry using off-axis parabolic mirrors. Room-temperature measurements were conducted under a dry $N_2$ atmosphere, whereas variable-temperature experiments were performed in a vacuum cryostat (pressure $< 1 \times 10^{-4}$ mbar). Time-resolved photoconductivity dynamics were obtained by fixing the electro-optic sampling point at the maximum of the THz transient and recording the pump-induced modulation of the THz field as a function of pump–probe delay. Frequency-resolved complex photoconductivity spectra were extracted from the full time-domain THz waveforms by Fourier transformation in the absence ($E(\omega)$) and in the presence of ($E'(\omega)$) photoexcitation. Under the thin-film approximation, the complex photoconductivity was calculated as:

$$\Delta\sigma(\omega) = -\frac{n_1+n_2}{Z_0 l}\left(\frac{E'(\omega)-E(\omega)}{E(\omega)}\right)$$

where $Z_0 = 377\,\Omega$ is the impedance of free space, $n_1$ and $n_2$ are the refractive indices of the incident and exit media, respectively, and $l$ is the sample thickness. The photoconductivity yield is calculated from the ratio of maximum photoconductivity and absorbed photon density. The temperature dependence of the carrier scattering rate $\gamma$ was analyzed using an Arrhenius model:

$$\gamma = B exp\left(-\frac{E_a}{k_B T}\right)$$

where $E_a$ is the activation energy, $B$ is the prefactor, and $k_B$ is the Boltzmann constant.

**Transient absorption (TA) spectroscopy.** Femtosecond TA measurements were performed using a Helios pump–probe spectrometer (Ultrafast Systems) driven by a regenerative-amplified Ti:sapphire laser (Coherent). The laser delivered ~25 fs pulses at 1.55 eV with a repetition rate of 1 kHz. The amplifier output was divided into two optical paths: one beam pumped an optical parametric amplifier (TOPAS-C) to generate 3.10 eV excitation pulses, while the other was focused into a sapphire plate to produce a broadband white-light continuum. The white-light output was split into probe and reference beams. The pump and probe pulses were spatially and temporally overlapped on the sample. A motorized delay stage controlled the temporal separation between the pump and probe pulses. The pump beam was modulated at 500 Hz using a mechanical chopper, and transient absorbance spectra were obtained by comparing the probe signals with and without photoexcitation.

**Computational details.** Geometry optimizations were performed using the Vienna Ab initio Simulation Package (VASP)[51-57] with PBE[58] functional and D3BJ[59] dispersion, using experimental lattice parameters. Band structure calculations were then performed using VASP with HSE06 hybrid functional[60]. Effective mass was then calculated from the band structure using Effmass package[61]. The exciton binding energy was estimated using the equation $E_b = E_{elec} - E_{opt}$. The optical band gap was obtained by performing Tauc analysis on the experimental UV-vis absorption data. Charge Density Difference (CDD) and its planar-averaged results along the stacking direction of the structures are calculated using VASP with PBE functional. VASPKIT code was used for post-processing of the results[62].

**Data availability**

All relevant data that support the findings are available within this article and its Supplementary Information. All quantum chemical calculations data are available in the NOMAD repository under the DOI: doi.org/10.17172/NOMAD/2026.03.26-1.

**Code availability**

All calculations presented in this work are performed using publicly available standard packages. All relevant information used for reproducibility can be found in the text and Supplementary Information.

**Methods-only references**

**Supplementary information**

Supplementary Figs. 1–12, Tables 1–3.

**Corresponding authors**

Correspondence to S.F., Z.W., M.B. or X.F.